# Parallel Image Thinning Through Topological Operators On Shared Memory Parallel Machines

Ramzi MAHMOUDI[1], Mohamed AKIL[1], Petr MATAS[1,2]

[1]Université Paris-Est, Laboratoire d'Informatique Gaspard-Monge, Equipe A3SI,
ESIEE Paris Cité Descartes, BP99, 93162 Noisy Le Grand, France

[2]Department of Applied Electronics and Telecommunications, University of West Bohemia,
Univerzitní 26, 306 14 Plzeň, Czech Republic

{mahmoudr, akilm, matasp}@esiee.fr

**Abstract**

*In this paper, we present a concurrent implementation of a powerful topological thinning operator. This operator is able to act directly over grayscale images without modifying their topology. We introduce an adapted parallelization methodology which combines split, distribute and merge (SDM) strategy and mixed parallelism techniques (data and thread parallelism). The introduced strategy allows efficient parallelization of a large class of topological operators including, mainly, λ-leveling, skeletonization and crest restoring algorithms. To achieve a good speedup, we cared about coordination of threads. Distributed work during thinning process is done by a variable number of threads. Tests on 2D grayscale image (512\*512), using shared memory parallel machine (SMPM) with 8 CPU cores (2× Xeon E5405 running at frequency of 2 GHz), showed an enhancement of 6.2 with a maximum achieved cadency of 125 images/s using 8 threads.*

## 1. Introduction

In many computer vision applications, standard techniques of pattern recognition are thinning algorithms. As a preprocessing stage, these algorithms have been used for the recognition of handwriting or printed characters, fingerprints, chromosomes and biological cell structures, etc. [1]. Topological thinning and skeletonization are ones of the most cardinal operators for this kind of preprocessing, especially since the development, by our team, of an efficient thinning algorithm able to act directly over grayscale image [2]. Using topological operators allows topology preservation which results in conservation of important significant information [3]. This conservation was impossible in the case of binary image processing [4]. Early thinning algorithms were designed for serial implementations, but since parallel computers are available several approaches have been developed with parallel processing [5,6]. In [5], Heydorn presents a concept for an implementation of different parallel thinning algorithms on parallel processors. The emphasis is put on a good parallelization using fine granularity and the simultaneous usage of vectorization.

This paper describes an adapted parallelization methodology combining split, distribute and merge (SDM) strategy based upon the well-known principle of divide-and-conquer and thread coordination which allows an efficient parallelism for introduced thinning operator on shared memory machines. Proposed strategy can also be applied for any topological operator having the same characteristics based on elementary operations of point characterization and similar algorithmic structure as we will demonstrate later.

This paper is organized as follows: in section 2, some basic notions of topological operators are summarized; the original algorithm of thinning is introduced. We define also the class of operators that our parallelization strategy may cover. In section 3, parallelization strategy, that has been adopted, is introduced. In section 4, Thread coordination and synchronization is discussed. In section 5, as a concrete example of introduced strategy and threads synchronization techniques, a parallel version of thinning algorithm is presented. Experimental analyzes results of different implementations are also presented and discussed. Finally, we conclude with summary and future work in section 6.

## 2. Topological thinning operator

Skeletonization and thinning are major applications of topology in image processing. A great number of thinning algorithms for binary images have been developed [7]. The use of this kind of images assumes a prior segmentation which implies a loss of information. Some attention has been given to the development of thinning algorithms acting directly over grayscale images. Dyer and Rosenfeld [8] proposed an algorithm based on a weighted notion of connectedness. The thinning is done directly over the graylevels of the points but, as the authors showed, the connectivity of objects is not always preserved. Other works [9] use an implicit image binarization into a background and a graylevel foreground. The graylevel information guides the removal of points of the foreground that are simple, in the binary sense. This technique makes it possible to obtain certain desired geometric properties. Inspired by this technique, M. Couprie and G. Bertrand [2] propose a filtered thinning method that allows to selectively simplify the topology, based on a local contrast parameter $\lambda$. To achieve this simplification; they introduce the notion of $\lambda$-destructible point which is more flexible then the notion of destructible point. This algorithm is the one we are going to present en details in the following to illustrate our parallelization strategy.

### 2.1. Theoretical background

First, we recall some basic notions of grayscale images. A 2D grayscale image may be seen as a map $F$ from $Z^2$ to $Z$. For each point $x \in Z^2$, $F(x)$ is the graylevel value of $x$. We denote by $\varphi$ the set composed by all maps from $Z^2$ to $Z$. Let $F \in \varphi$, the section of F at the level $k$ is the set $F_k$ composed of all point $x \in Z^2$ such that $F_k \geq k$. As for the binary case, if we use the $n$-adjacency for the section $F_k$ of F , we must use $\overline{n}$-adjacency for the section $\overline{F_k}$ with $(n, \overline{n}) =$ (8,4) or (4,8). We remind that for two points $x(x_1, x_2)$, $y(y_1, y_2) \in Z^2$, we consider that $y$ is 4-adjacent to $x$ if $|y_1 - x_1| + |y_2 - x_2| \leq 1$, and $y$ is 8-adjacent to $x$ if $\max(|y_1 - x_1|, |y_2 - x_2|) \leq 1$. In the following, we consider the two neighborhoods relations $\Gamma_4$ and $\Gamma_8$ defined by, for each point $x \in Z^2$,
$\Gamma_4(x) = \{y \in Z^2 \mid y \text{ is } 4 \text{ adjacent } to \ x\}$,
$\Gamma_8(x) = \{y \in Z^2 \mid y \text{ is } 8 \text{ adjacent } to \ x\}$.

For more general presentation, we will define $\Gamma_n^*(x) = \Gamma_n(x)\backslash\{x\}$. We will also denote by $\overline{F}$ the complementary map of F. We note that the complementary sets of the section of F are section of $\overline{F}$.

In all the rest of this paragraph, we will note n=8 for the section of F , thus we must use $\overline{n}$=4 for $\overline{F}$. It is also important to mention that a non-empty connected component $X$ of a section $F_k$ of $F$ is a (regional) maximum for $F$ if $X \cap F_{k+1} = \emptyset$ and a set $X \subset Z^2$ is a regional minimum for $F$ if it is a regional maximum for $\overline{F}$. Let $F \in \varphi$, the point $x \in Z^2$ is destructible (for F) if x is a simple for $F_k$, with $k = F(x)$. We remind that a point x is said simple for $X \subset Z^2$ if $T(x, X) = 1$ and $\overline{T}(x, X) = 0$ with $T(x, X)$ and $\overline{T}(x, X)$ the two connectivity numbers defined as follows (# $X$ stands for the cardinal of $X$): $T(x, X) = \#C_n[x, \Gamma_8^*(x) \cap X]$; $\overline{T}(x, X) = \#C_{\overline{n}}[x, \Gamma_8^*(x) \cap \overline{X}]$; So we can define the four neighborhoods:

$\Gamma^{++}(x, F) = \{y \in \Gamma_8^*(x); F(y) > F(x)\}$
$\Gamma^{+}(x, F) = \{y \in \Gamma_8^*(x); F(y) \geq F(x)\}$
$\Gamma^{--}(x, F) = \{y \in \Gamma_8^*(x); F(y) < F(x)\}$
$\alpha^-(x, F) = \begin{cases} \max\{F(y), y \in \Gamma^{--}(x, F), if \Gamma^{--}(x, F) \neq \emptyset\} \\ F(x) \qquad\qquad\qquad\qquad\qquad\qquad otherwise \end{cases}$

We define also some associated connectivity numbers:
$T^+(x, F) = \#C_n[x, \Gamma^+(x, F)]$
$T^{++}(x, F) = \#C_n[x, \Gamma^{++}(x, F)]$
$T^{--}(x, F) = \#C_n[x, \Gamma^{--}(x, F)]$

Furthermore, the connectivity numbers allow the classification of the topological characteristics of a point:
$x$ is a <u>peak point</u> if $T^+(x, F) = 0$.
$x$ is a <u>k-divergent</u> if $T^{--}(x, F) = k \text{ with } k > 1$.

A point is said to be a $\lambda$-deletable point (for F), $\lambda$ being a positive integer, if it is either a $\lambda$-destructible point, or a peak point such that $F(x) - \alpha^-(x, F) \leq \lambda$. We remind that a point x is said $\lambda$-destructible if it satisfies one of the two following conditions: x is destructible or x is k-divergent and at least k-1 connected components $c_i$ of $\Gamma^-(x, F)$ are such that $F(x) - F^-(c_i) \leq \lambda$, with $i = \{1, \dots, k-1\}$.
Let $X \subset Z^2$ and $x \in X$, x is an end point (for X) if $\#(\Gamma_n^*(x) \cap X)=1$. Let $F \subset \varphi$ and $x \in Z^2$, x is an end point (for F) if it is an end point for the set $F_k$ with $k = F(x)$. A point is said to be $\lambda$-end point (for F) if it is an end point for F and if: $F(x) - \alpha^-(x, F) > \lambda$.

### 2.2. Original algorithm

$\forall F \in \varphi$, we say that $G \in \varphi$ is a skeleton of F if G is obtained from F by iteratively selecting a destructible and non-end point in F and lowering it down to $\alpha^-(x, F)$, until stability. In order to get a filtered skeleton, that is to eliminate non significant branches and regional minima, Bertrand and Couprie allow $\lambda$- deletable and not λ-end to be lowered. It is important to mention that each time that a pixel is lowered, its eight neighbors must be reexamined to be sure that topology is still preserved. In Figure 1, we illustrate this method on a gradient image (a) obtained from a 2D grayscale image of an MRI brain section by Deriche gradient operator. (b) is obtained by a filtered thinning with $\lambda = 10$.

| **Algorithm** : λ –Skeleton (input : F ∈ $\varphi$, λ ∈ N; output : F) |
|---|
| 1. Repeat until stability |
| 2.    Among all the points which are λ–deletable and not λ–end |
| 3.       Select a point x of minimal value ; |
| 4.       $F(x) \coloneqq \alpha^-(x, F)$; |

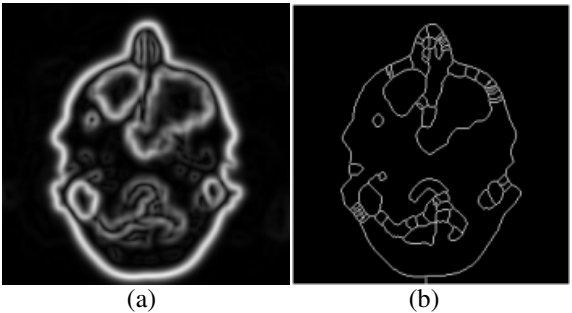

Fig. 1. (a): after Deriche gradient operator; (b) filtered skeleton with $\lambda = 10$.

### 2.3. Class of operators based upon point characterization in the grayscale image case

Bertrand [1,10] introduced connectivity numbers for grayscale image as showed in section 2.1. These numbers describe locally (in a neighborhood of 3∗3) the topology of a point. According to this description any point can be characterized following its topological characteristics. He also introduced some elementary operations able to modify gray level of a point without modifying image topology. These elementary operations of point characterization present the fundamental link of large class of topological operators including, mainly, skeletonization and crest restoring algorithms [2]. This class can also be extended, under condition, to homotopic kernel and leveling kernel transformation [11], topological 2D and 3D object smoothing algorithm [12] and topological watershed based on w-thinning algorithm [13]. All mentioned algorithms get also many algorithmic structure similarities. In fact associated characterizations procedures evolve until stability with induce common recursivety between different algorithms. Also the grey level of any point can be lowered or enhanced more than once. Finally, all the mentioned algorithms get a pixel's array as input and output. It is important to mention that, to date, this class has not been efficiently parallelized like other classes as connected filter of morphological operator which recently has been parallelized in Wilkinson's work [14].

### 3. Parallelization strategies

Multiprocessor chips make computing more efficient by exploiting parallelism which is one of the outstanding challenges of modern computer sciences. Exploiting such parallelism depends on the way of scheduling tasks to different processors such that the tasks can be computed simultaneously in parallel [15]. Computing each individual task in parallel using all the processors and computing tasks one after the other is Data Parallelism [16]. For both strategies, programming challenges arise at all scales of multiprocessor systems: at the small scale, processors within a single chip need to coordinate access to shared memory locations; at the large scale, processors in a super computer need to coordinate routing of data. It is also possible to combine the mentioned strategies for better scheduling [17]; such strategy is called Mixed Parallelism. In this case, challenges are also related to the asynchronous criteria of modern computers: activities can be halted without warning by interrupts, preemption or frequently by cache misses.

In a more global frame, better strategies taking advantage of such parallelism to improve computational speed are based on the well known principle of divide and conquer. The application of this principle cannot be independent from the type of algorithm [18]. Indeed, application of this principle to divide the initial problem and then application of Mixed Parallelism strategy during parallel sub-problems processing seems to be sufficient. But it is only true for static parallel algorithms in which each thread can achieve its work "independently" from the other. Low-level image processing algorithms are a good example of this class because they have a high degree of locality allowing different segments of the image to be treated independently by different processors [19]. Other global operators like Fourier transform and Euclidean distance transform are separable, allowing parallelization by defining a direction for computing pixels [20].

For target algorithms, as we shown in section 2.3, get some iterative criteria and evolutes until stability. Intermediate results need also to be stored. Each time, that a pixel is lowered, a new process for inserting its neighborhoods is launched. So threads need imperatively to communicate and to share the same queue; this is why we return to dynamically parallel algorithms in which threads can interact with one another. Through parallelization strategies presented in the beginning of this section, we see that for an inter-processor parallelism based on divide and conquer principle; better performance can be achieved by the use of mixed parallelism, since it allows us to combine SDM-Strategy and coordination of threads. And as our processing continues until stability, we primarily focus on an approach where data parallelism is used at upper levels. At lower levels of the processing, we will switch to threads parallelism and coordination to compute parallel read/write for managing cache-resident data. If we observe carefully the studied class, we see that there are two fundamental stages: the first one is to characterize a point. Then, according to the nature of this point, we decide to eliminate it (modify its value) or not. If one pixel is lowered, it becomes necessary to re-examine its eight neighbors. So we can follow these steps to apply divide and conquer principle for our class of algorithms.

### 3.1. How to Split

In upper level, search space is subdivided into smaller regions, and bounds are found on all solutions contained in each sub-region under consideration. Usually, dividing original image is not advised, when dealing with topological operators, warning topology is not preserved. But pixel characterization procedure can be split into sub-procedures. So we can characterize in parallel more than one point during a single iteration.

### 3.2. How to distribute

All algorithms associated to our class are executed in a loop until stability for example: no more λ–deletable and not λ–end for thinning. Thus we can specify translation states. The initial system will undergo an evolution until reaching a stable state. During the evolution, sub-procedures defined in section 3.1 are distributed among used threads. Usually thread evolution is uniform, but, due to data dependency, thread evolution must be dynamic. The number of threads is changing during the whole processing procedure. So the second stage of the algorithm can be realized. If a point is characterized, its value is lowered and its eight neighbors will be inserted in a FIFO queue. Since one thread terminates, it will generate a new thread to repeat the same work with new inserted neighbors. This is how we can plan distribution.

### 3.3. How to merge

The key problem of each parallelization is merging obtained results. Normally this phase is done at the end of the process when all results are returned by all threads what usually means that only one output variable is declared and shared between all threads. But as we mentioned in section 3.2, we are dealing with a dynamic evolution so we can plan the following: since two threads finished, they directly merge and a new thread is created. This implies the creation of some shared FIFO queue containing all inserted neighbors by both two parent threads. Only one shared data structure will contain pixels lowered by all threads. In threads merging, there is no hierarchical order, the only criteria is finish time. It is also important to mention that only newly created threads can modify the created FIFO queue and one neighbor cannot be inserted twice. It is a precaution in order to minimize consumed cache.

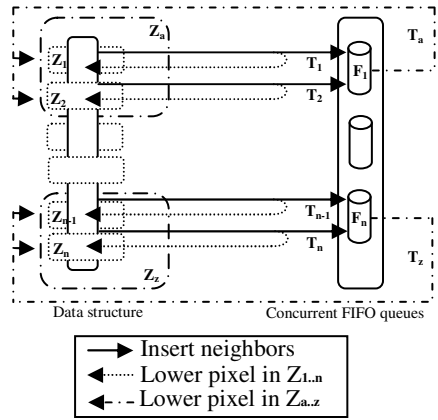

Fig. 2: Merging of threads and associated area of activities.

In Fig. 2, we illustrate the introduced SDM-strategy. The original shared data structure, containing all pixels, is divided in n research zones $\{z_1, z_2, .. z_{n-1}, z_n\}$. We associate one thread from the following list $\{T_1, T_2, .. T_{n-1}, T_n\}$ to each zone. Each thread can browse freely its zone and if it detects target types, it lowers the characterized pixel and it pushes its eight neighbors in one of the shared concurrent FIFO queues. One queue cannot be shared by more than two threads. There is no hierarchical order in merging of threads. Queues are attributed for the first two threads which have arrived (first-come, first-served). Since two threads finish their work, a new thread is created to browse their FIFO queue. For storing new value of possibly found target point, the new thread has full access to both original zones. Example, thread $T_a$ merging from $T_1$ and $T_2$ got access to $Z_a$ with $Z_a = Z_1 + Z_2$.

## 4. Coordination of threads

Here is the second major challenge in multi-core multithread architecture programming. In an ideal case, moving from one-core to multi-core should provide n-fold increase in computational power. But practically, it is something that never happened. In fact, all existing computational problems cannot be efficiently parallelized without incurring the costs of inter-processor coordination. Let's come back to our algorithm, consider eight threads which cross eight search spaces in order to characterize pixels then push its eight neighbors in a FIFO queue.

This kind of analysis was evoked in many researches. Let's focus on Amdahl's Law [21]. It captures the notion that the extent to which we can speed up any complex work is limited by percentage of the sequential part in the executed work. Definition of the speedup S of a work is the ratio between the time it takes *one* processor to complete the work versus the time it takes $n$ concurrent processors to complete the same work. Amdahl's Law defines the maximum speedup $S$ that can be achieved by $n$ processors collaborating on an application, where $p$ is the fraction of the work that can be executed in parallel. Assume, for simplicity, that a single processor completes the work in one second. With $n$ concurrent processors, the parallel part takes $(p/n)$ seconds and the sequential part takes $(1-p)$ seconds. Overall, the parallelized computation takes $(1 - p + \frac{p}{n})$ seconds.

So the speedup is: $(n) = \frac{1}{1-p+\frac{p}{n}}$.

Through this formula, for the given problem and an eight-core machine, Amdahl's law says that even if we parallelize 90% of the solution, but not the remaining 10%, then we end up with only four-fold speedup, and not the expected eight-fold speedup. In fact, these additional parallel parts involve substantial communication and coordination.

In our dynamic parallelization strategy, as we explained in section 3.3, each two threads will share only one FIFO queue in order to push neighbors of lowered pixels. Intuitively we are going to opt towards a solution with a simple lock-based shared FIFO queue. Associated push and pop methods will be synchronized by a mutual exclusion lock. Even if this implementation is a correct concurrent FIFO queue, because each method accesses and updates fields while holding an exclusive lock, the method calls take effect sequentially. And according to Amdahl's law, this sequential communication can substantially affect the performance of our program as a whole. In multi-core architecture, such synchronization technique can also be the origin of costly overheads.

Even if we opt to second method based on lock-free solution [22] in order to minimize the overheads, it is demanded that at least one thread (of all the threads that are executing the push or pop function at one moment) is progressing (inserting or extracting pixels from or to the FIFO queue). Unfortunately, we do not know in advance how many parallel threads will call push or pop functions. And method calls still take effect sequentially. Other solution is wait-free technique [23], it is required that a process finishes within a finite number of execution steps. Something that we cannot also guarantee because we cannot predict how many points will be characterized and then how many pixels will be inserted in the FIFO queue.

Finally we decide to move to spin-wait mechanism [24], for illustration we propose figure 4, a thread waiting to push an item might spin for a brief duration without being added to the queue of waiting threads. As a result, the thread is effectively put to sleep without relinquishing the remainder of its CPU time slot. It is potentially more efficient to spin and wait, instead of using either lock-free or wait-free mechanisms, because those force a thread context switch, which is one of the most expensive operations performed by the operating system.

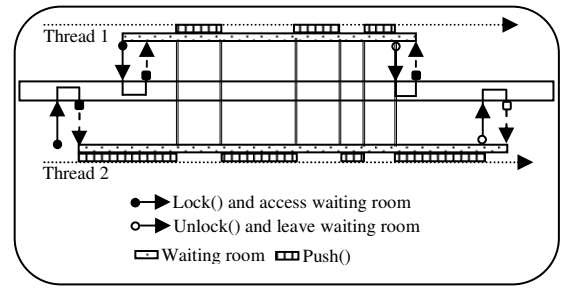

Fig. 4: Spin-wait Synchronization

## 5. Performance testing

### 5.1. Parallel λ -Skeleton algorithm

Now we present a parallel version of the thinning according to the concepts previously discussed. Let the map F from $Z^2$ to $Z$ represent the input grayscale image. For each point $x \in Z^2$, F(x) is the graylevel value of x. We denote by φ the set composed by all maps from $Z^2$ to $Z$. Let $F \in φ$, the section of F at the level k is the set $F_k$ composed of all point $x \in Z^2$ such that $F_k \geq k$. Let $T$ be the set of type sought in the characterization of pixels. For thinning algorithm: T = {λ– deletable and not λ– end points}. It is important to mention that points from *T* can also be end-point and isolate-point for crest restoring. We will refer to global search space by *Ime*, and associated map (from $Z^2$ to Z ) to each sub-space $Ime_i$ is $F_i$ . For each point x ∈ $Z^2$, $F_i(x)$ is the graylevel value of x in the search space $Ime_i$. The following dynamically parallel λ–Skeleton algorithm (it is adapted for two concurrent threads, but it can be easily extended to N threads) starts by dividing the search space. $m_{inf}$ and $m_{sup}$ define sub-region bounds. Since the distributed work starts, each thread will lower each characterized pixel and then push its eight neighbors in $E_{sn}$. $E_{sn}$ is the set of all selected neighbors and it is shared between only two threads. $E_{sn}$ will be the newly defined set to explore since the threads finished. Newly characterized pixels are pushed in a private set called $E_{ki}$. The pixel set assigned to the newly generated thread is nothing else than $E_{sn}$ and the associated search space is $((Ime_i ∪ Ime_{i+1}) ∪ E_{ki} ∪ E_{ki+1})$.

**Algorithm** :Dynamically Parallel λ –Skeleton
1. $For\ all\ p ∈ Ime\ do$
2. $if\ (m_{inf} < Ime_i(p) < m_{sup})then\ E_i ← E_i ∪ \{p\};$
3. $Repeat\ until\ stability$
4. $E_{sn} ← ∅;$
5. $While\ (k ≠ 0)then$
6. $For\ all\ p ∈ E_i\ do$
7. $if\ (p\ ∈ T)\ then\ F_i(x) ← α^-(x, F_i(x));$
8. $E_{sn} ← E_{sn} ∪ \{eight\ p\ neighbors\};$
9. $else\ E_{ki} ← E_{ki} ∪ \{p\};$
10. $endif$
11. $For\ all\ p ∈ E_{i+1}\ do$
12. $if\ (p\ ∈ T)\ then\ Ime_{i+1}(x) ← α^-(x, Ime_{i+1}(x));$
13. $E_{sn} ← E_{sn} ∪ \{eight\ p\ neighbors\};$
14. $else\ E_{ki+1} ← E_{ki+1} ∪ \{p\};$
15. $endif$
16. $E_i ← E_{sn};$
17. $Ime_i ← Ime_i + Ime_{i+1};$
18. $Ime_i ← Ime_i ∪ E_{ki} ∪ E_{ki+1};$
19. $if\ (E_i = ∅)\ then\ k ← 0;$
20. $clean\ \{E_i, E_{ki}, E_{ki+1}\};$
21. $end\ while$

### 5.2. Experimental analyses

The proposed parallel λ -Skeleton algorithm was implemented in C in two variants: the first implementation, based on a simple lock-based shared FIFO queue, using OpenMP critical directive. The second is based on a spin-wait FIFO queue. Wall-clock execution times for numbers of threads equal to 1, 2, 4, 8, and 16, for each one of these implementations, were determined. The efficiency measure $Ψ(n)$ is given by the following formula With *n* the number of processors:

$$Ψ(n) = sequential\ time/(n ∗ parallel\ time)$$

Times were performed on eight-core (2× Xeon E5405) shared memory parallel computer of the Faculty of Electrical Engineering and Communication of Brno University, on Intel Quad-core Xeon E5335, on Intel Core 2 Duo E8400 and Intel mono-processor Pentium 4 660. Each processor of the Xeon E5405 and E5335 runs at 2 GHz and both of the two machines have 4 GB of RAM. The E8400 processor runs at 3 GHz. The Pentium processor runs at 3.6 GHz (see Table 1). The last two machines have 2 GB of RAM. The minimum value of 5 timings was taken as most indicative of the speed of the algorithm. The measurements were done on 2D grayscale image (512*512) of real brain MRI. Results of the two implementations are shown in Figure 5 and Figure 7.

|  | P4 660 | E8400 | E5335 | E5405 |
|---|---|---|---|---|
| CPU Speed | 3.6 GHz | 3 GHz | 2 GHz | 2 GHz |
| Bus Speed | 800 MHz | 1333 MHz | 1333 MHz | 1333 MHz |
| L2 Size | 4 MB | 6 MB | 8 MB | 12 MB |
| L2 Speed | 3.6 GHz | 3 GHz | 2 GHz | 2 GHz |

Table 1. Characteristics of processors

On the eight-core machine, wall-clock execution time for the first implementation using a lock-based shared FIFO queue drops from an average of 40.211 ms for a single thread down to 28.458 ms at 8 threads. For the second implementation using spin-wait FIFO queue, wall-clock execution time drops from an average of 41.889 ms for a single thread down to 8.282 ms at 8 threads. As expected, the speed-up for the second implementation using Private-Shared FIFO queue is higher than for the one using lock-based shared FIFO queue, because context changing were nearly eliminated.

A remarkable result shown in figure 6 and figure 8 is the fact that the speed-up increases as we increase the number of threads beyond the number of processors in our machine (eight cores). For the first implementation, the speedup at 8 threads is 1.7 ± 0.05. However, for the second implementation the speedup has increased to 6.2 ± 0.01. Another common result between figure 5 and figure 7 is stability of execution time on each *n*-core machine since the code uses *n* or more threads.

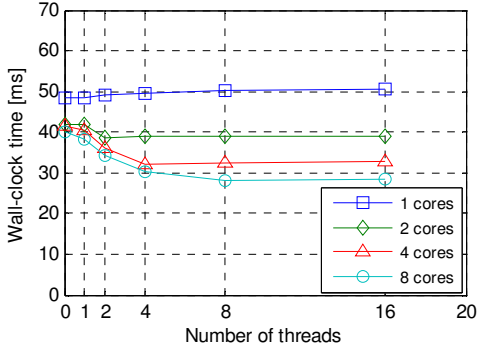

Fig. 5: wall-clock execution time for the first implementation using a lock-based shared FIFO queue.

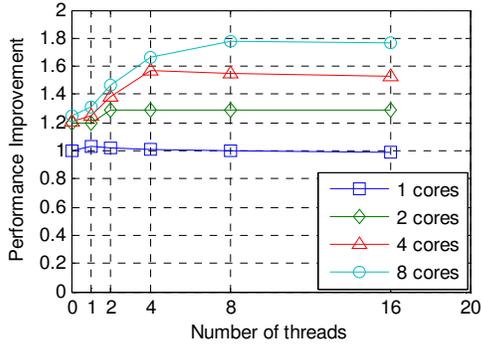

Fig. 6: Performance improvement for the first implementation using a lock-based shared FIFO queue.

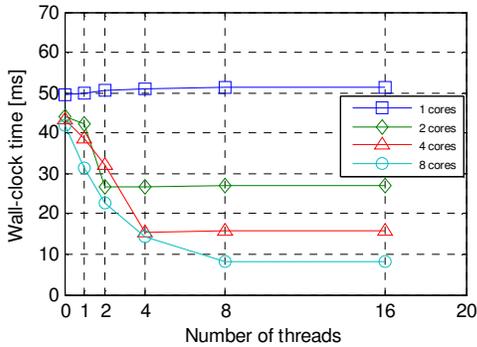

Fig. 7: wall-clock execution time for the second implementation using a spin-wait shared FIFO queue.

For better readability of our results, we tested the efficiency of our algorithm on various architectures using the $\Psi(n)$ formula introduced earlier with fixed serial time equal to 48.247 ms. For parallel time we use best parallel time obtained using 8 threads. As can be seen in Figure 9, second implementation is more efficient that the first one in all architectures. It is also suitable to return to Amdahl's law, introduced in section 4, in order to explain obtained results. In fact the global speed up formula is $S(n) = \frac{T(1)}{T(n)}$. Then the defined efficiency $\Psi(n) = T_s/(n * T_p)$ can be written

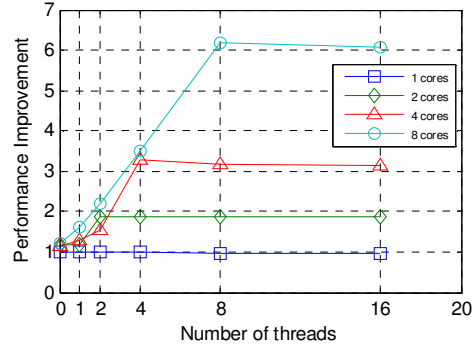

Fig. 8: Performance improvement for the second implementation using a spin-wait FIFO queue.

as $\Psi(n) = T_s/(n * T_p) = \frac{S(n)}{n} = \frac{T(1)}{n * T(n)}$. According to Amdahl's law $S(n) = \frac{1}{1-p+\frac{p}{n}}$, efficiency can be written as follows: $\Psi(n) = \frac{1}{n*(1-p)+p}$

Thus if the number of cores increases, the speedup also increases (more work can be done simultaneously with more threads). On the other hand the efficiency will decrease.

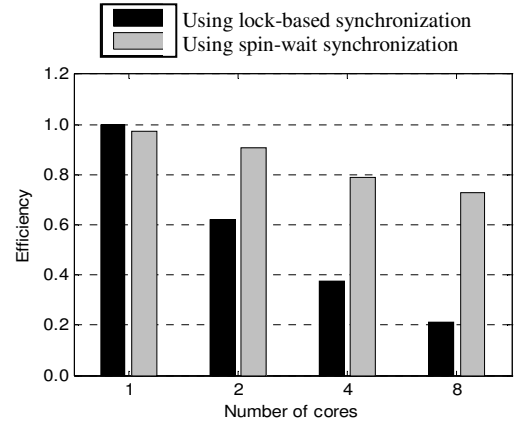

Fig. 9: Efficiency improvement

## 6. Conclusion

In this paper, we have presented a new parallel version of the λ–Skeleton algorithm. We have also presented a adapted parallelization strategy combining Split Distribute and Merge (SDM) strategy and mixed parallelism techniques. SDM-strategy was a conditional application of the well known principle of divide and conquer. Associated mixed parallelization techniques were data parallelism at upper levels and thread parallelism at lower levels of the processing.

The first major contribution in this paper is the non-specific nature of the proposed parallelization strategy. In fact, the introduced strategy can be applied to a large class of topological operators introduced in section 2.3.

The second contribution concerned threads parallelism and more specifically threads coordination and communication during computing dependently parallel read/write for managing cache-resident data which present a substantial problem. The problem addressed by this paper is how to deal with shared FIFO queue which requires inter-process coordination and communication in an essential way. And thanks to combination of spinning and waiting techniques, the proposed algorithm shows a good degree of speed-up using eight threads (about 6.2 on eight cores of the 2× Xeon E5405, about 3.1 on the Quad-cores of Xeon E5335 and 1.8 on Core 2Duo E8400).

Parallel topological operator computation poses many challenges, ranging from parallelization strategies to coding and implementation techniques. We tackle these challenges using successive refinement, starting with highly local operators, which process only by characterizing points and then deleting target pixels, and gradually moving to more complex topological operators with non-local behavior. In future work, we will study parallel computation of the topological watershed [25].

## 6. References

[1] Anil K. Jain, Fundamentals of Digital Image Processing, Prentice-Hall, United States, Oct. 1988.

[2] M. Couprie, F. N. Bezerra, and G. Bertrand, "Topological operators for grayscale image processing", Journal of Electronic Imaging Vol. 10, Oct. 2001, pp. 1003-1015.

[3] J. C. Everat, and G. Bertrand, "New topological operators for segmentation", International Conference On Image Processing, Vol. 3, Sep. 1996, pp. 45-48.

[4] T. Ojala, M. Pietikäinen, and T. Mäenpää, "Generalized local binary pattern operators for multi-resolution grayscale and rotation invariant texture classification", Advances In Pattern Recognition Conf. Brazil Vol. 2013, March 2001, pp. 397-406.

[5] S. Heydorn, and P. Weidner, "Optimization and Performance Analysis of Thinning Algorithms on Parallel Computers" Parallel Computing 17.1991, pp. 17-27.v

[6] N. BOURBAKIS, N. STEFFENSEN, and B. SAHA, "Design of an array processor for parallel skeletonization of images", IEEE transactions on circuits and systems. 2, Analog and digital signal processing, 1997, pp. 284-298

[7] L. Lam, L. Seong-Whan, and C. Y. Suen, "Thinning methodologies: a comprehensive survey", IEEE transactions on pattern analysis and machine intelligence vol. 14, 1992, pp. 869-885.

[8] R. C. Dyer, and A. Rosenfeld, "Thinning Algorithms for Gray-Scale Pictures", Pattern Analysis and Machine Intelligence, IEEE Transactions Vol. PAMI-1, Jan. 1979, page(s): 88-89

[9] Y. Shiaw-Shian, and T. Wen-Hsiang, "A new thinning algorithm for gray-scale images by the relaxation technique", Pattern recognition vol. 23, 1990, pp. 1067-1076.

[10] G. Bertrand, J. C. Everat and M. Couprie, "Topological approach to image segmentation", In SPIE Vision Geometry V, vol. 2826, 1996, pp. 65-76.

[11] G. Bertrand, J. C. Everat, and M. Couprie, "Image segmentation through operators based on topology," Journal of Electronic Imaging, 1997, pp. 395-405.

[12] M. Couprie, and G Bertrand, "Topology preserving alternating sequential filter for smoothing 2D and 3D objects" Journal of Electronic Imaging, Vol. 13, 2004, pp. 720-730.

[13] G. Bertrand, "On Topological Watersheds", Journal of Mathematical Imaging and Vision, Vol. 22, 2005, pp. 217 – 230.

[14] M.H.F. Wilkinson, H. Gao, W.H. Hesselink, J.-E. Jonker and A. Meijster, "Concurrent Computation of Attribute Filters on Shared Memory Parallel Machines", Pattern Analysis and Machine Intelligence, IEEE Transactions on, Oct. 2008, pp. 1800-1813.

[15] T. G. Mattson, B. A. Sanders and B. L. Massingill, "Patterns for Parallel Programming" Addison-Wesley Professional, First Edition, Sep. 2004, Chapter 4.

[16] P. J. Hatcher, and J. M. Quinn, "Data-Parallel Programming on MIMD Computers" Mit Press, Des. 2001, Chapter 1.

[17] S. Feldmann, J. Sgall, and S-H Teng, "Dynamic scheduling on parallel machines", 32nd Annual Symposium on Foundations of Computer Science, 1.Oct 1991, pp. 111-120.

[18] L. Wangqing, S. Mingren, and P. Ogunbona, "A New Divide and Conquer Algorithm for Graph-based Image and Video Segmentation" Multimedia Signal Processing, 2005, pp.1-4.

[19] F. J. Seinstra, D. Koelma, and J. M. Geusebroek, "A software architecture for user transparent parallel image processings," Parallel Computing, vol. 28, 2002, pp. 967-99.

[20] A. Meijster, J. Roerdink, and W. Hesselink, "A general algorithm for computing distance transforms in linear time," in Proc. Int. Symp. Math. Morphology (ISMM) 2000, pp. 331–340.

[21] G. M. Amdahl, "Validity of the single-processor approach to achieving large scale computing capabilities" AFIPS Conference Proceeding vol. 30, Atlantic City, NJ., 1967, pp. 483-485.

[22] J. H. Anderson, S. Ramamurthy and k. Jeffay, "Real Time computing with lock-free shared Object" ACM Transaction on Computer System Vol.15,1997, pp.134 -165.

[23] M. Herlihy, "Wait-Free Synchronization", ACM transaction on programming languages and system, vol. 13, 1991, pp. 124 - 149

[24] T. E. Anderson, "The performance of spin lock alternatives for shared-money multiprocessors", IEEE Transactions on Parallel and Distributed Systems, Vol. 1, Jan. 1990, pp:6 -16.

[25] J. Cousty, M. Couprie, L. Najman and Gilles Bertrand "Weighted fusion graphs: Merging properties and watersheds". Discrete Applied Mathematics 156, 2008, pp. 3011-3027.